\documentclass[12pt]{article}
\usepackage{graphicx}

\catcode`\@=11 \@addtoreset{equation}{section}\catcode`\@=12
\newcommand{\be}{\begin{equation}}
\newcommand{\ee}{\end{equation}}
\newcommand{\ba}{\begin{eqnarray}}
\newcommand{\ea}{\end{eqnarray}}

\newcommand{\fnm}{\footnotemark}
\newcommand{\fnt}{\footnotetext}

\begin{document}
 \thispagestyle{empty}

 \vskip 35mm

 \begin{center}
 {\large\bf  Billiard representation for pseudo-Euclidean
 Toda-like systems \\ of cosmological origin}

  \vspace{15pt}

 \normalsize\bf
 V.D. Ivashchuk\fnm[1]\fnt[1]{ivashchuk@mail.ru}
 and  V.N. Melnikov\fnm[2]\fnt[2]{melnikov@phys.msu.ru}
¨
 \vspace{5pt}

 \it Center for Gravitation and Fundamental Metrology,
 VNIIMS, 46 Ozyornaya ul., Moscow 119361, Russia  \\

 Institute of Gravitation and Cosmology,
 Peoples' Friendship University of Russia,
 6 Miklukho-Maklaya ul.,  Moscow 117198, Russia \\


 \end{center}
 \vskip 10mm

The pseudo-Euclidean Toda-like system of cosmological origin is
considered. When certain restrictions on the parameters of the
model are imposed, the dynamics of the model near the
``singularity'' is reduced to a billiard on the
$(n-1)$-dimensional Lobachevsky space $H^{n-1}$. The geometrical
criterion for the finiteness of the billiard volume and its
compactness is suggested. This criterion reduces the problem to
the problem of illumination of $(n-2)$-dimensional sphere
$S^{n-2}$ by point-like sources. Some examples are considered.

 \vskip 10mm


 \vskip 30mm

 \pagebreak


\section{Introduction}
\setcounter{equation}{0}

In this paper we consider pseudo-Euclidean Toda-like system
described by the following Lagrangian
 \begin{equation} \label{1.1}
 L = {L}(z^{a}, \dot{z}^{a}, {\cal N}) = \frac{1}{2} {\cal N}^{-1}
 \eta_{ab} \dot{z}^{a} \dot{z}^{b} -  {\cal N} {V}(z),
 \end{equation}
where ${\cal N} >0 $ is the Lagrange multiplier (modified lapse function),
 \\
 $(\eta_{ab})=  diag(-1,+1, \ldots ,+1)$  is matrix of minisuperspace
metric, $a,b = 0, \ldots , n-1$, and
\begin{equation} \label{1.2}
{V}(z) = \sum_{\alpha=1}^{m} A_{\alpha} \exp(u^{\alpha}_a z^a)
\end{equation}
 is the potential. Here $A_{\alpha} \neq 0$.

Models of such type occur in multidimensional cosmology (see, for
example, \cite{1}-\cite{6} and references therein) as well as in
multicomponent 4-dimensional cosmology \cite{7}-\cite{10}. The
Lagrange systems (\ref{1.1}), (\ref{1.2}) are not well studied
yet. We note, that the Euclidean Toda-like systems are more or
less well studied \cite{12}-\cite{18} (at least for certain sets
of parameters, associated with finite-dimensional Lie algebras or
affine Lie algebras). There is also a criterion of integrability
by quadrature (algebraic integrability) for the Euclidean
Toda-like systems established by Adler and van Moerbeke \cite{16}.
We note that in gravitational context the Euclidean Toda-like
systems were first considered in \cite{19}-\cite{21}.

In this paper we consider the behavior of the dynamical system
(\ref{1.1}) for $n \geq 3$ in the limit
 \begin{equation} \label{1.3}
 z^2  \equiv  -(z^0)^2 + (\vec{z})^2   \rightarrow  -\infty, \qquad
 z =(z^0, \vec{z}) \in {\cal V}_{-}, \end{equation}

where ${\cal V}_{-} \equiv \{(z^0, \vec{z}) \in R^n | z^0 < -
|\vec{z}| \}$ is the lower light cone. The limit (1.3) implies

 \be  \label{1.4}
 z^0 \rightarrow  -\infty
 \ee

and under certain additional assumptions describes the approaching
to the singularity in corresponding cosmological models. We impose
the following restrictions on the vectors $u^{\alpha} =
 (u^{\alpha}_0, \vec{u}^{\alpha})$ in the potential (\ref{1.2})

 \begin{eqnarray} \label{1.5}
 && 1) \ A_{\alpha} > 0, \ {\rm if} \ (u^{\alpha})^2 =
 -(u^{\alpha}_0)^2 + (\vec{u}^{\alpha})^2 > 0;
 \\ \label{1.6}
 && 2) \ u^{\alpha}_0 > 0  \ {\rm for \ all} \  \alpha  = 1,
 \ldots, m.
 \end{eqnarray}

We note that in multidimensional cosmology  a special interest is
connected with the investigations  of oscillatory behavior of
scale factors near the singularity \cite{22}-\cite{26}. This
direction in higher-dimensional gravity was stimulated by
well-known results for
 "mixmaster" model \cite{7}-\cite{10}. We note, that there is also an elegant
explanation for oscillatory behavior of scale factors of
Bianchi-IX model suggested by Chitre \cite{9,10} and recently
considered in \cite{27}-\cite{29}. In the Chitre's approach the
Bianchi-IX cosmology near the singularity is reduced to a billiard
on the Lobachevsky space $H^2$ (see Fig. 4 below). The volume of
this billiard is finite. This fact together with the well-known
behavior (exponential divergences) of geodesics on the spaces of
negative curvature may lead to a stochastic behavior of the
dynamical system in the considered regime \cite{30,31}.

Here we consider the generalization of Chitre's approach \cite{9}
to the multidimensional case \cite{26}.  In the limit (\ref{1.3})
the dynamics of the model (\ref{1.1}), (\ref{1.2}) (with the
restrictions (\ref{1.5}), (\ref{1.6}) imposed) is reduced to a
billiard on the $(n-1)$-dimensional Lobachevsky space $H^{n-1}$
 (Sec. 2). The geometrical criterion for the finiteness of the
billiard volume and its compactness is suggested.  This criterion
reduces the considered problem to the geometrical (or topological)
problem of illumination of $(n-2)$-dimensional  unit sphere
 $S^{n-2}$ by $m_{+} \leq m$ point-like sources located outside the
sphere \cite{32,33}.  These sources correspond to the components
with  $(u^{\alpha})^2 > 0$ (Sec. 2). When these sources illuminate
the sphere then, and only then, the billiard has a finite volume
and the corresponding cosmological model possesses an oscillatory
behavior near the singularity. In this case from topological
requirements we obtain the restriction on the number of components
with $(u^{(\alpha)})^2  > 0$:  $m_{+} \geq n$, i.e. this number
should be no less than the minisuperspace dimension. In Sec. 3 we
illustrate the suggested approach using the  examples of
Bianchi-IX  and prototype cosmological models.

It should be noted that fixing the gauge in (\ref{1.1}) ${\cal N}
  =1$ we get Lagrange system
 \begin{equation} \label{1.7}
 L_1 = \frac{1}{2} \eta_{ab} \dot{z}^{a} \dot{z}^{b} -  {V}(z),
 \end{equation}
with the zero-energy constraint
 \be \label{1.8}
 E_1 = \frac{1}{2} \eta_{ab} \dot{z}^{a}
 \dot{z}^{b} + {V}(z) = 0.
 \ee

For non-zero energies $E_1$ we are lead to Lagrange systems with
the  Lagrangians

 \be \label{1.9}
 L_2 = L_1 + \frac{1}{2} \varepsilon
 (\dot{z}^{n})^2
 \ee

 $\varepsilon = - {\rm sign} E_1 $, and  the zero-energy
constraint $E_2 = 0$.  Thus, for $E_1 < 0 $ we get the
pseudo-Euclidean Toda-like system (\ref{1.1}) with $n$ replaced by
 $n+1$ and the potential (\ref{1.2}).  In cosmology this corresponds to
addition of scalar field.  For $E_1 >0$ we obtain the Lagrangian
system with the kinetic term of signature $(-,+, \ldots, +,-)$. In
cosmology this corresponds to the incorporation of a phantom
field. We also note that the Euclidean Toda-like systems may be
embedded into (\ref{1.1}) by considering the potentials
 (\ref{1.2}) with  $u^{\alpha}_0 = 0$.


\section{Billiard  representation}
\setcounter{equation}{0}

Here we consider the behavior of the dynamical system, described
by the Lagrangian (\ref{1.1}) for $n \geq 3$ in the limit
 (\ref{1.3}). We restrict the Lagrange system (\ref{1.1}) on ${\cal
  V}_{-}$, i.e. we consider the Lagrangian
 \begin{equation} \label{2.1}
  L_{-} \equiv L|_{TM_{-}} , \qquad M_{-} = {\cal V}_{-} \times R_{+},
 \end{equation}
where $ TM_{-}$ is tangent vector bundle over $M_{-}$  and
 $R_{+} \equiv \{ {\cal N} > 0 \}$. (Here $F|_{A}$ means the restriction
of function $F$ on $A$.)
Introducing  an analogue of the Misner-Chitre coordinates in
 $\cal{V}_{-}$ \cite{9,10}
 \begin{eqnarray} \label{2.2}
 &&z^0 = - \exp(-y^0) \frac{1 + \vec{y}^2}{1 - \vec{y}^2},
 \\ \label{2.3}
 &&\vec{z} = - 2 \exp(-y^0) \frac{ \vec{y}}{1 - \vec{y}^2},
 \end{eqnarray}
$|\vec{y}| < 1$, we get for the Lagrangian (1.1)

 \begin{equation} \label{2.4}
 L_{-} = \frac{1}{2} {\cal N}^{-1} e^{- 2 y^0}
 [- (\dot{y}^{0})^2 + {h_{ij}}(\vec{y}) \dot{y}^{i} \dot{y}^{j}]
 -  {\cal N} V.
 \end{equation}

Here
 \be \label{2.5}
 {h_{ij}}(\vec{y}) = 4 \delta_{ij} (1 - \vec{y}^2)^{-2},
 \ee
 $i,j =1, \ldots , n-1$, and

 \be \label{2.6}
 V = {V}(y) =
 \sum_{\alpha=1}^{m} A_{\alpha} \exp {\bar{\Phi}}(y,u^{\alpha}),
 \end{equation}

where
 \be \label{2.7}
 {\bar{\Phi}}(y,u)  \equiv - e^{-y^0}(1 - \vec{y}^2)^{-1}
 [u_0 (1 + \vec{y}^2) + 2 \vec{u}\vec{y}],
 \ee

We note that the $(n-1)$-dimensional open unit disk (ball)
 \be \label{2.8}
 D^{n-1} \equiv \{ \vec{y}= (y^1, \ldots, y^{n-1})| |\vec{y}| < 1 \}
 \subset R^{n-1}
 \ee

with the metric $h = {h_{ij}}(\vec{y}) dy^i \otimes dy^j $ is one
of the realization of the $(n-1)$-dimensional Lobachevsky space
$H^{n-1}$.

We fix the gauge
 \be \label{2.9}
  {\cal N} =   \exp(- 2y^0) = - z^2.
 \ee
Then, it is not difficult to verify that the Lagrange equations
for the Lagrangian (\ref{1.1}) with the gauge fixing (\ref{2.9})
are equivalent  to  the Lagrange equations for the Lagrangian
 \be \label{2.10}
 L_{*} = - \frac{1}{2}  (\dot{y}^{0})^2 +  \frac{1}{2}
 {h_{ij}}(\vec{y}) \dot{y}^{i} \dot{y}^{j} -  V_{*}
 \ee
with the energy constraint imposed
 \be \label{2.11}
 E_{*} = - \frac{1}{2}  (\dot{y}^{0})^2 +  \frac{1}{2}
 {h_{ij}}(\vec{y}) \dot{y}^{i} \dot{y}^{j} +  V_{*} = 0.
 \ee
 Here
 \be \label{2.12}
 V_{*} =  e^{-2y^0} V =
 \sum_{\alpha=1}^{m} A_{\alpha} \exp({\Phi}(y,u^{\alpha})),
 \ee
where
 \be \label{2.13}
 {\Phi}(y,u)  = - 2y^0 + {\bar{\Phi}}(y,u).
 \ee

Now we are interested in the behavior of the dynamical system in
the limit  $y^0 \rightarrow - \infty$ (or, equivalently, in the
limit (\ref{1.3}) implying (\ref{1.4}). Using the relations ($u_0
 \neq 0$ )
 \begin{eqnarray} \label{2.14}
 &&{\Phi}(y,u) = - u_0 \exp(-y^0)
 \frac{{A}(\vec{y}, -\vec{u}/u_0)}{1 - \vec{y}^2} - 2y^0,
 \\\label{2.15}
  && {A}(\vec{y}, \vec{v}) \equiv
  (\vec{y} - \vec{v})^2 -\vec{v}^2 + 1,
  \end{eqnarray}
 we get
 \be \label{2.16}
 \lim_{y^0 \rightarrow - \infty} \exp {\Phi}(y,u)  = 0
 \ee
 for $u^2 = - u_0^2 + (\vec{u})^2 \leq 0$, $u_0 > 0$ and
 \be \label{2.17}
 \lim_{y^0 \rightarrow - \infty} \exp {\Phi}(y,u)  =
 {\theta_{\infty}}(-{A}(\vec{y}, - \vec{u}/u_0))
 \ee
for $u^2 > 0$, $u_0 > 0$. In (2.17) we denote

 \ba \label{2.18}
 {\theta_{\infty}}(x) \equiv + &\infty, &x \geq 0,  \nonumber \\
                              & 0    , &x < 0.
 \ea
Using restrictions (\ref{1.5}), (\ref{1.6}) and relations
 (\ref{2.12}), (\ref{2.16}), (\ref{2.17}) we obtain

 \be \label{2.19}
 {V_{\infty}}(\vec{y}) \equiv \lim_{y^0
 \rightarrow - \infty} {V_{*}}(y^0, \vec{y}) = \sum_{\alpha \in
 \Delta_{+}} {\theta_{\infty}}(-{A}(\vec{y}, -
 \vec{u^{\alpha}}/u_0^{\alpha})). \ee

Here we denote
 \be \label{2.20}
 \Delta_{+}  \equiv \{ \alpha |
 (u^{\alpha})^2 > 0 \}.
 \ee

The potential $V_{\infty}$
may be also written as following
 \ba \label{2.21}
 {V_{\infty}}(\vec{y}) =
 {V}(\vec{y},B) \equiv &0, &\vec{y} \in B,
 \nonumber \\
 &+ \infty, &\vec{y}
 \in D^{n-1} \setminus B,
 \ea
where

\be \label{2.22}
 B = \bigcap_{\alpha \in
 \Delta_{+}} {B}(u^{\alpha})  \subset D^{n-1}, \ee

 \be \label{2.23}
 {B}(u^{\alpha})  = \left\{ \vec{y} \in D^{n-1} :
 \left|\vec{y} + \frac{\vec{u}^{\alpha}}{u_{0}^{\alpha}}\right| >
 \sqrt{\left(\frac{\vec{u}^{\alpha}}{u_0^{\alpha}}\right)^2 - 1} \right\},
 \ee

 $\alpha \in \Delta_{+}$. $B$ is an open domain. Its boundary
 $\partial B = \bar{B} \setminus B$ is formed by certain parts of
 $m_{+} = |\Delta_{+}|$ ($m_{+}$ is the number of elements in
 $\Delta_{+}$) of $(n-2)$-dimensional spheres with the centers in
the points

\be \label{2.24}
 \vec{v}^{\alpha} = -
 \vec{u}^{\alpha}/u^{\alpha}_{0}, \qquad \alpha \in \Delta_{+},
 \ee

($|\vec{v^{\alpha}}| > 1$) and radii

 \be \label{2.25}
  r_{\alpha} = \sqrt{(\vec{v}^{\alpha})^2 - 1}
 \ee
respectively (for $n =3$, $m_{+} = 1$, see Fig. 1).

\setcounter{figure}{0}

\begin{figure}[htbp]\center
\includegraphics[width=90mm]{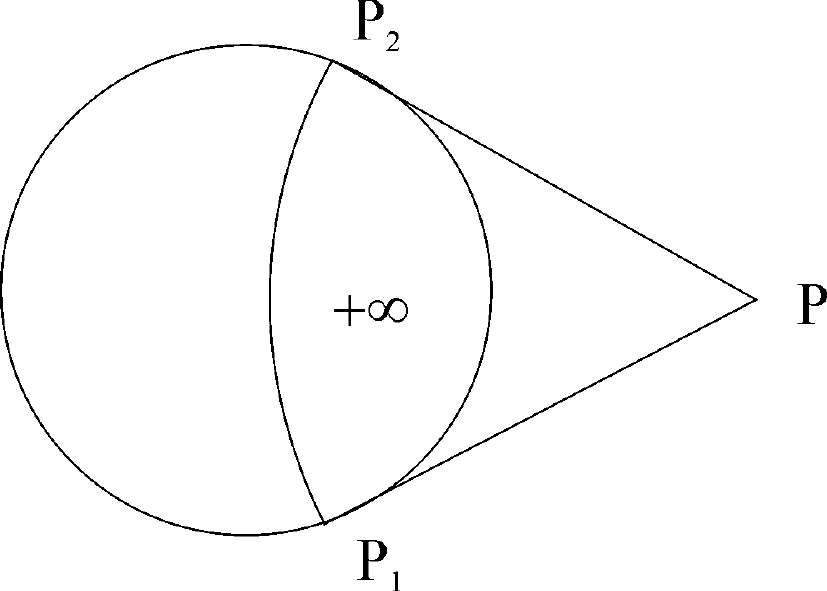}
\caption{An example of billiard for $n=3$, $m_{+} = 1$.}\label{F5}
\end{figure}

So, in the limit $y^{0} \rightarrow - \infty$ we are led to the
dynamical system
\ba
       \label{2.26}
 &L_{\infty} = - \frac{1}{2} (\dot{y}^{0})^2 +  \frac{1}{2}
 {h_{ij}}(\vec{y}) \dot{y}^{i} \dot{y}^{j} -  {V_{\infty}}(\vec{y}), \\
 \label{2.27}
 &E_{\infty} = - \frac{1}{2} (\dot{y}^{0})^2 +  \frac{1}{2}
 {h_{ij}}(\vec{y}) \dot{y}^{i} \dot{y}^{j} +  {V_{\infty}}(\vec{y}) = 0,
 \ea
which after the separating of $y^0$ variable

\be
   \label{2.28}
 y^0 = \omega (t - t_0),
 \ee

 ($\omega \neq 0$ , $t_0$  are constants) is reduced to the
Lagrange system with the Lagrangian
 \be \label{2.29}
  L_{B} =  \frac{1}{2} {h_{ij}}(\vec{y})
 \dot{y}^{i} \dot{y}^{j} -  {V}(\vec{y},B). \ee

 Due to (2.28)
 \be \label{2.30}
 E_{B} =  \frac{1}{2}
 {h_{ij}}(\vec{y}) \dot{y}^{i} \dot{y}^{j} +  {V}(\vec{y},B) =
 \frac{\omega^2}{2}.
 \ee

We put $\omega > 0$, then the limit $t \rightarrow - \infty$
corresponds to (\ref{1.3}).  When the set (\ref{2.20}) is empty
 ($\Delta_{+} = \emptyset$), we have $B = D^{n-1}$ and the
Lagrangian (\ref{2.29}) describes  the  geodesic  flow  on the
Lobachevsky space $H^{n-1} = (D^{n-1}, h_{ij} dy^i \otimes dy^j)$.
In this case there are two families of non-trivial geodesic
solutions (i.e. ${y}(t) \neq const$):
 \ba \label{2.31}
 1. &{\vec{y}}(t) = \vec{n}_1
 [\sqrt{v^2 -1} \cos {\varphi}(t) - v] +  \vec{n}_2 \sqrt{v^2 -1} \sin
                 {\varphi}(t) ,  \\
 \label{2.32}
 &{\varphi}(\bar{t})  = 2 \arctan [(v -
   \sqrt{v^2 -1}) \tanh (\frac{1}{2} \omega (t-t_1)], \\
  \label{2.33}
   2.  &{\vec{y}}(t)
  = \vec{n}_2 \tanh (\frac{1}{2} \omega (t-t_1)).
 \ea
Here $\vec{n}_1^2 = \vec{n}_2^2 =1$, $\vec{n}_1 \vec{n}_2 = 0$,
 $v > 1$, $\omega > 0$,  $t_1 = const$.

Graphically the first solution corresponds to the arc of the circle with
the center at point ($-v \vec{n}_1$) and the radius $\sqrt{v^2 -1}$.
This circle belongs to the plane spanned by vectors
  $\vec{n}_1$ and  $\vec{n}_2$
(the centers of the circle and the ball $D^{n-1}$ also belong to
this plane). We note, that the solution (\ref{2.31})-(\ref{2.32})
in the limit $v \rightarrow \infty$ coincides with the solution
 (\ref{2.33}).

When  $\Delta_{+} \neq \emptyset$ the Lagrangian
 (\ref{2.33}) describes the motion of the particle  of  unit  mass,  moving
in the ($n-1$)-dimensional generalized billiard $B \subset
 D^{n-1}$  (see (\ref{2.22}).  The geodesic motion in $B$ (\ref{2.31})-(\ref{2.33})
corresponds to a "Kasner epoch" and the reflection from the
boundary corresponds to the change of Kasner epochs.  For $n = 3$
some examples  of (2-dimensional) billiards are depicted in Figs.
 2-4.

\begin{figure}[htbp]\center
\includegraphics[width=90mm]{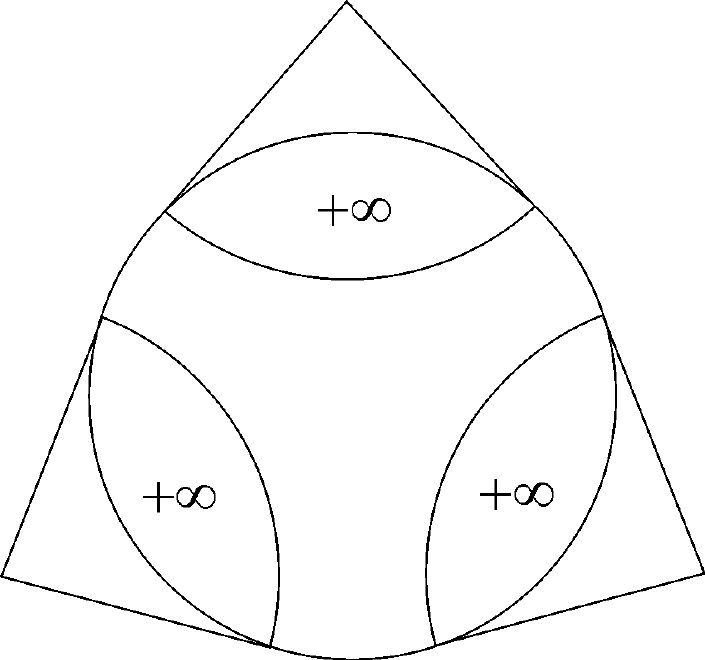}
\caption{Billiard with infinite volume for $n=3$, $m_{+} = 3$.}\label{F6}
\end{figure}

\begin{figure}[htbp]\center
\includegraphics[width=90mm]{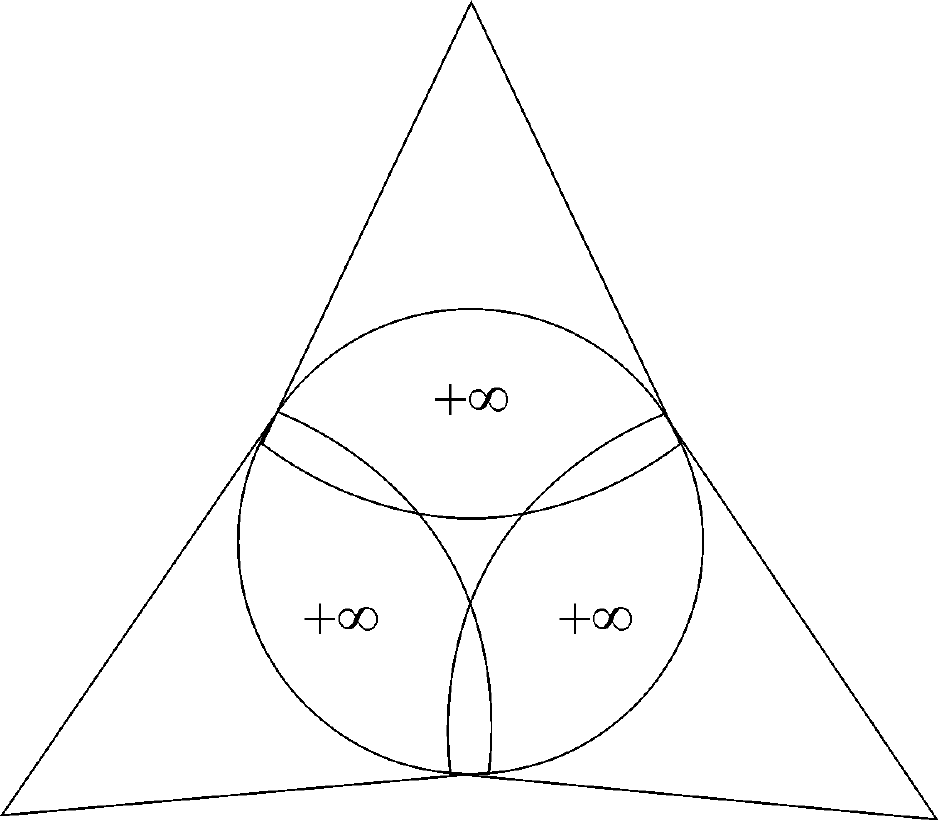}
\caption{Compact billiard for $n=3$, $m_{+} = 3$.}\label{F7}
\end{figure}

\begin{figure}[htbp]\center
\includegraphics[width=90mm]{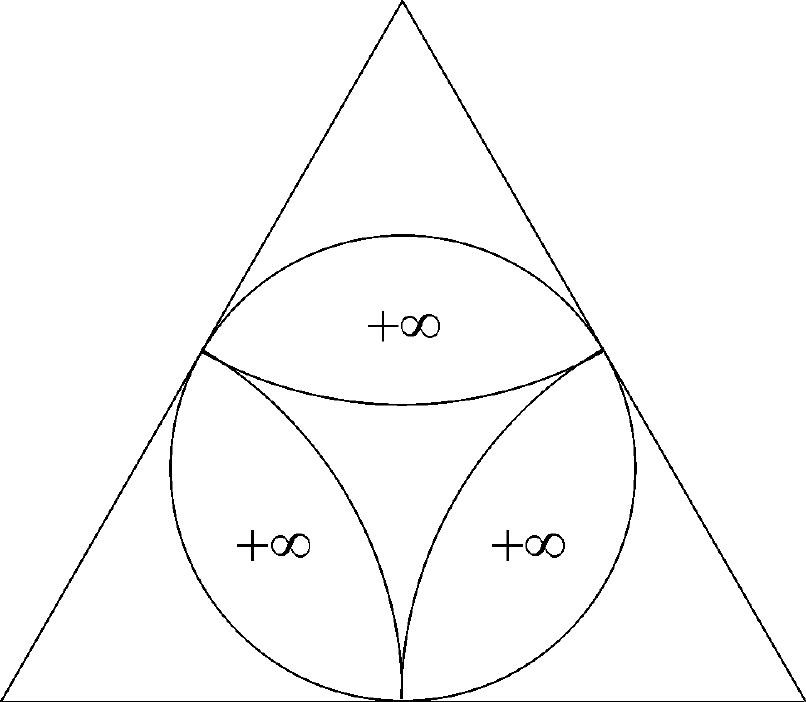}
\caption{Billiard corresponding to Bianchi-IX model
(non-compact with finite volume).}\label{F8}
\end{figure}

We note, that the boundary of the billiard $\partial B$ is formed
by geodesics. For some billiards  this fact  may be used for
``gluing'' certain parts of boundaries.

The billiard $B$ in Fig. 2. has an infinite volume: $vol B =
+\infty$. In this case there are three open zones  at  the
infinite  circle $|\vec{y}| =1$. After a finite number of
reflections from the boundary the  particle (in general position)
moves  toward  one  of  these  open zones. For corresponding
cosmological model we get the "Kasner-like" behavior in the limit
 $t \rightarrow - \infty$ \cite{4}.

For billiards depicted in Figs. 3 and 4 we have $vol B < +
\infty$. In the first case (Fig. 3) the closure of the billiard
$\bar{B}$ (in the topology of $D^{n-1}$) is compact (in the
topology of $D^{n-1}$) and in the second case (Fig. 4) $\bar{B}$
is non-compact. In these two cases we have an ``oscillatory-like''
motion of the particle.

Analogous arguments may be applied  to the case $n > 3$.
So, we are interested  in  the configurations with finite volume of $B$.

We propose a simple geometric criterion for the finiteness of the
volume of $B$ and compactness of $\bar{B}$ in terms of the
positions of the points (\ref{2.24})  with respect to the
($n-2$)-dimensional unit sphere $S^{n-2}$ ($n \geq 3$).

{\bf Proposition 1}. The billiard $B$ (\ref{2.22}) has a finite
volume if and only if the point-like sources of light located at
the points $\vec{v}^{\alpha}$ (\ref{2.24}) illuminate the unit
sphere $S^{n-2}$. The closure of the billiard  $\bar{B}$ is
compact (in the topology of $D^{n-1} \simeq H^{n-1}$) if and only
if the sources at points (\ref{2.24}) strongly illuminate
 $S^{n-2}$.

{\bf Definition}. Here the point is called strongly illuminated if
it belongs to the interior (open) part of the illuminated region.
At Fig. 1 the source $P$ strongly illuminates the open arc $(P_1,P_2)$.

{\bf Proof}. We consider the set  $\partial^c B  \equiv B^c \setminus
 \bar{B}$, where $B^{c}$ is the completion of $B$ (or, equivalently, the
closure of $B$ in the topology of $R^{n-1}$). We recall that $\bar{B}$
is the closure of $B$ in the topology of $D^{n-1}$. Clearly, that
 $\partial^{c} B$ is a closed subset of $S^{n-2}$, consisting of all those
points that are not strongly illuminated by sources (\ref{2.24}).
There are three possibilities:  i) $\partial^c B$ is empty; ii)
 $\partial^{c} B$ contains some interior point (i.e. the point
belonging to $\partial^{c} B$ with some open neighborhood); iii)
 $\partial^{c} B$ is non-empty finite set, i.e. $\partial^{c} B  =
 \{ \vec{y}_1, \ldots \vec{y}_l \}$. The first case i) takes place
if and only if $\bar{B}$ is  compact in the topology of $D^{n-1}$.
Only in this case the sphere $S^{n-2}$ is strongly illuminated by
the sources (\ref{2.24}). Thus the second part of proposition is
proved. In the case i) $vol B$ is finite. For the volume we have

 \be \label{2.34}
 vol B = \int_{B} d^{n-1} \vec{y}
 \sqrt{h} = \int_{0}^1 dr (1- r^2)^{1-n} S_{r}.  \ee

The "area" $S_{r} \rightarrow C > 0$ as $r \rightarrow 1$ in the
case ii) and, hence, the integral (2.34) is divergent.  In the
case iii)

 \be \label{2.35}
 S_{r} \sim C_1
 (1 - r)^{2(n-2)}  \ as  \   r \rightarrow 1
 \ee

 ($C_1 >0$) and, so, the
integral (\ref{2.34}) is  convergent.  Indeed, in the case iii),
when  $r \rightarrow 1$,  the "area" $S_{r}$ is the sum of $l$
terms. Each of these terms is the $(n-2)$-dimensional "area" of a
transverse side of a deformed pyramid with a top at some point
 $\vec{y}_k$, $k =1, \ldots , l$. This multidimensional pyramid is
formed by certain parts of spheres orthogonal to $S^{n-2}$ in the
point of their intersection $\vec{y}_k$. Hence, all lengths of the
transverse section  $r = const$ of the "pyramid" behaves like $(1
  - r)^{2}$, when  $r \rightarrow 1$,  that justifies (\ref{2.35}). But
the unit sphere $S^{n-2}$ is illuminated by the sources
 (\ref{2.24}) only in the cases i) and iii). This completes the
proof.

To our knowledge, the problem of illumination of convex body in a
vector space by point-like sources for the first time was
considered in \cite{32,33}. For the case of $S^{n-2}$ this problem
is equivalent to the problem of covering the spheres with spheres
 \cite{34,35}. There exists a topological bound on the number of
point-like sources $m_{+}$ illuminating the sphere $S^{n-2}$
 \cite{33}:
 \be \label{2.36}
  m_{+} \geq n.
 \ee
Thus, we are led to the following proposition.

{\bf Proposition 2.} When $m_{+} < n$, i.e. the number of
components with $(u^{\alpha})^2 > 0$ is less than the
 minisuperspace dimension, the billiard $B$ (\ref{2.22}) has
infinite volume: $vol B = +\infty$.

In this case there exist an open zones on the sphere $S^{n-2}$ and
the oscillatory  behaviour near the singularity for cosmological
models is absent (we get a Kasner-like behaviour for $t
 \rightarrow - \infty$ \cite{4,38}).

Remark 1. Let the points (\ref{2.24}) form an open convex
polyhedron $P \subset R^{n-1}$. Then the sources at (\ref{2.24})
illuminate $S^{n-2}$, if $D^{n-1} \subset P$, and strongly
illuminate $S^{n-2}$, if $\overline{D^{n-1}} \subset P$.

 \section{Some examples}
 \setcounter{equation}{0}

 \subsection{Bianchi-IX cosmology}

Here we  consider the well-known ``mixmaster'' model \cite{7,8}
with the metric

 \begin{equation}  \label{3.1}
 g_{mix} \equiv - \exp[2{\gamma}(t)] dt \otimes dt +
 \sum_{i=1}^{3} \exp[2{x^{i}}(t)] e^i \otimes e^i,
 \end{equation}

where 1-forms  $e^i = {e^i_{\nu}}(\zeta) d \zeta^{\nu}$ satisfy
the relations
 \be \label{3.2}
 de^i = \frac{1}{2} \varepsilon_{ijk} e^j \wedge e^k,
 \ee

  $i,j,k = 1,2,3$. The forms $e^i$ are defined on $S^3 \simeq {SU}(2)$
and are the components of the Morera-Cartan form (see,for example
 \cite{36}) on  ${SU}(2)$. The Einstein equations for the metric (\ref{3.1})
are equivalent to the Lagrange equations for the Lagrangian

 \begin{equation} \label{3.3}
 L = {L}(\gamma, x, \dot{x}) =
 \frac{1}{2} {\cal N}^{-1} G_{ij}
 \dot{x}^{i}\dot{x}^{j}- {\cal N} {V}(x).
 \end{equation}
Here

 \begin{equation} \label{3.4}
 G_{ij}=\delta_{ij}- 1
 \end{equation}

are the components of the minisuperspace metric,

 \be \label{3.5}
 V = V_{mix} \equiv   \frac{1}{4}
 (e^{4x^1} + e^{4x^2} + e^{4x^3}
 - 2 e^{2 x^1 + 2 x^2} - 2 e^{2 x^2 + 2 x^3} - 2 e^{2 x^1 + 2 x^3})
 \ee
is the potential and the Lagrange multiplier is

 \begin{equation} \label{3.6}
 {\cal N} = \exp( \gamma - {\gamma_{0}}(x)) > 0,
 \ee

where

 \be \label{3.7}
 \gamma_{0} = \sum_{i=1}^{n} x^{i}
 \ee

 and $n = 3$.

In $z$-variables  [1,2]

 \ba \label{3.8}
 &&z^0 =  q^{-1} \sum_{i=1}^{n} x^i, \qquad q = [n/(n-1)]^{1/2},
 \\ \label{3.9}
 &&z^a = [1/ (n - a + 1)(n -a )]^{1/2} \sum_{j=a+1}^{n} (x^j - x^a),
 \ea

 $a = 1, \ldots ,n-1$, where here $n = 3$, we get the Lagrangian
 (\ref{1.1}) with 3-vectors

 \be \label{3.10}
 u^{1} = \frac{4}{\sqrt{6}} (1,1, - \sqrt{3}), \ u^{2} =
 \frac{4}{\sqrt{6}} (1,1, + \sqrt{3}), \ u^{3} = \frac{4}{\sqrt{6}}
 (1,-2,0),
 \ee

\be  \label{3.11}
 u^{4} = \frac{1}{2}(u^{1} + u^{2}), \ u^{5} = \frac{1}{2}(u^{1} + u^{3}), \
 u^{6} = \frac{1}{2}(u^{2} + u^{3}),
 \ee

and, consequently,
 \be \label{3.12}
 (u^{\alpha})^2  = 8, \qquad
 (u^{3 + \alpha})^2 = 0,
 \ee

 $\alpha = 1,2,3$. Thus, the conditions (\ref{1.5}), (\ref{1.6}) are satisfied.
The components with $\alpha = 4,5,6$ do not "survive" in the
approaching to the singularity (see (\ref{2.16})). For the vectors
 (\ref{2.24}) we have

 \be \label{3.13}
 \vec{v}^1 = (1, - \sqrt{3}), \ \vec{v}^2 = (1, + \sqrt{3}),  \
 \vec{v}^3 = (-2, 0),
 \ee
i.e. a triangle from Fig. 4 (see also \cite{27}). In this case the
circle
 $S^1$ is illuminated by sources located at points $\vec{v}^i$,
 $i= 1,2,3$, but is not strongly illuminated. In agreement with
Proposition 1 the billiard $B$ has a finite volume, but $\bar{B}$
is not compact.

\subsection{Prototype model}

Now, we consider the model, described by Lagrangian (\ref{3.3})
with minisupermetric (\ref{3.4}), $i,j = 1, \ldots n$,  and the
potential
 \be
  V = \frac{1}{4} \sum_{(i,j,k) \in S}
 \exp[2 \sum_{i=1}^{n} x^i + 2(x^i - x^j - x^k)]
 \ee
where $S = \{ (i,j,k)|
 i,j,k = 1, \ldots, n;  j \neq i \neq k, \, j < k \}$,
 $n > 2$. This model may be considered as a prototype cosmological
model describing the behaviour of the solutions to the Einstein
equations in the dimension $D = 1+ n$  near the singularity
 \cite{22,23}. The sum in (\ref{3.14}) contains $n(n-1)(n-2)/2$ terms.
In $z$-coordinates (\ref{3.8}), (\ref{3.9}) we get the Lagrangian
 (\ref{1.1}) with $(u^{\alpha})^2 = 8 > 0$ and $u^{\alpha}_0 = 2/q
  > 0$ for any component $\alpha \in S$. Thus, the restrictions (\ref{1.5})
and (\ref{1.6}) are satisfied.

The corresponding billiard is not compact for all dimensions,
and has a finite volume for $n < 10$ and  infinite volume for $n \geq 10$
 \cite{38}. This proposition follows from Proposition 1 formulated in terms
of inequalities on Kasner parameters \cite{38} and analysis of
this inequalities performed in Ref. \cite{23}. For  $n=4$ the
billiard was studied recently in \cite{37}.

 \subsection{Scalar field generalization}

Let us consider the Lagrangian (\ref{1.1}) with
 $a,b = 0, \ldots , n$ and  $u^{\alpha}_{n} = 0$, i.e.
the potential does not depend upon the "scalar field" $\varphi =
 z^n$. In this case the correponding billiard has an infinite
volume, since at least  two points $(0, \ldots, 0, \pm 1)$ ("North
and South Poles") on the "Kasner sphere" $S^{n-1}$ are not
illuminated by the sources at the points $\vec{v}^{\alpha}$  with
 $v^{\alpha}_n = 0$. Thus the oscillatory behaviour in this case is
absent. We may expect that the points belonging to the domain of
shadow on the  sphere $S^{n-1}$ are integrals of motion in this
case. Due to (\ref{1.9}) this analysis may be also applied to
pseudo-Euclidean Lagrange systems (\ref{1.7}) in the region of
negative energy $E_1 < 0$.

 \section{Discussions}
 \setcounter{equation}{0}

Thus, we obtained the "billiard  representation" for
pseudo-Euclidean Toda-like system (\ref{1.1}), (\ref{1.2})
 and proved the  geometrical criterion
for the finiteness of the billiard volume and the compactness of the billiard
(Proposition 1, Sec. 2). This criterion may be used as a rather effective
(and universal) tool for the selection of the cosmological models with an
oscillatory behavior near the singularity.

 Remark 2. It may be shown  that the condition (1.5) may be weakened by
 the following one
 \be \label{4.1}
 u^{\alpha}_0 > 0, \  if  \ (u^{\alpha})^2 \leq 0.
 \ee

In this case there exists a certain generalization of the set
 ${B}(u^{\alpha})$ from (\ref{2.23}) for arbitrary $u^{\alpha}_0$
 ($(u^{\alpha})^2 > 0$).  The Proposition 1 (Sec. 2) should be
modified by including into consideration the sources at infinity
(for $u^{\alpha}_0 = 0$) and "anti-sources" (for $u^{\alpha}_0 <
 0$). For "anti-source" the shadowed domain coincides with the
illuminated domain for the usual source (with $u^{\alpha}_0 > 0$).
In this case we deal with  the  kinematics of tachyons
 ($\vec{v}^{\alpha}$ are the velocities of tachions).

\begin{center} {\bf Acknowledgments}   \end{center}

The authors would like to thank our colleagues K.A. Bronnikov,
 A.A. Kirillov, \\
 M.Yu. Konstantinov, A.G. Radynov, P. Spindel and the participants
of the Marcel Grossmann meeting (Stanford, 1994) for useful
discussions. We are  grateful to R.V. Galiullin for pointing out
the attention to ref. \cite{33}.

This work was supported in part by the Russian Ministry of Science.


\pagebreak

\end{document}